\documentclass[12pt,oneside]{JHEP3}
\usepackage{graphics}
\usepackage{amsfonts}
\usepackage[latin1]{inputenc}
\usepackage{epsfig}
\usepackage{fancyhdr}
\usepackage{cite}

\title{A Comment on Quantum Distribution Functions  
and the OSV Conjecture} 
\author{César Gómez$^{1,2}$\phantom{x} and\phantom{x} Sergio Monta\~nez$^1$\\ {\it ${}^1\!$ Instituto de F\'\i 
sica Te\'orica CSIC/UAM,}\\ {\it C-XVI Universidad Aut\'onoma,}\\ {\it E-28049 Madrid \rm 
Spain}\\
{\it ${}^2\!$ Theory Group, Physics Department, CERN,}\\
{\it CH-1211 Geneva 23, \rm Switzerland}
\\
E-mail: 
\email{cesar.gomez@uam.es}, 
\email{sergio.montannez@uam.es} }

\abstract{Using the attractor mechanism and the relation between the quantization of $H^{3}(M)$ and  
topological strings on a Calabi Yau threefold $M$ we define a map from BPS black holes into coherent states. This map allows us to  
represent the Bekenstein-Hawking-Wald entropy as a quantum  
distribution function on the phase space $H^{3}(M)$. This  
distribution function is a mixed Husimi/anti-Husimi distribution  
corresponding to the different normal ordering prescriptions for the  
string coupling and deviations of the complex structure moduli. From the integral representation  
of this distribution function in terms of the Wigner distribution we  
recover the Ooguri-Strominger-Vafa (OSV) conjecture in the region ``at infinity" of the complex structure 
moduli space. The physical meaning of the OSV corrections are briefly  
discussed in this limit.
}

\preprint{
CERN-PH-TH/2006-169\\
IFT-UAM/CSIC-06/39\\
{\tt hep-th/0608162}
}


\newcommand{\ba}{\begin{array}} 
\newcommand{\ea}{\end{array}} 
\newcommand{\be}{\begin{equation}} 
\newcommand{\ee}{\end{equation}}
\newcommand{\ben}{\begin{equation}} 
\newcommand{\een}{\end{equation}} 
\newcommand{\bea}{\begin{eqnarray}} 
\newcommand{\eea}{\end{eqnarray}} 
\newcommand{\gsim}{\mathrel{\mathop{\kern 0pt \rlap 
  {\raise.2ex\hbox{$>$}}} \lower.9ex\hbox{\kern-.190em $\sim$}}}

\def\entero{\mathbb{Z}}
\def\real{\mathbb{R}}
\def\complejo{\mathbb{C}}
\def\imaginario{\mathbb{I}}

\def\I{\mbox{Im}}
\def\R{\mbox{Re}}
\def\pd{\partial} 
\def\a{\alpha} 
\def\bet{\beta} 

\def\g{\gamma} 
\def\d{\delta}

\def\t{\tau} 
\def\l{\lambda}

\newcommand{\Ene}{\ensuremath{{\cal{N}}}} 
\newcommand{\Ce}{\ensuremath{{\cal{C}}}}
\newcommand{\Eme}{\ensuremath{{\cal{M}}}}
\newcommand{\Efe}{\ensuremath{{\cal{F}}}}
\newcommand{\Ele}{\ensuremath{{\cal{L}}}}
\newcommand{\De}{\ensuremath{{\cal{D}}}}
\newcommand{\Hache}{\ensuremath{{\cal{H}}}}

\newcommand{\ra}{\rangle}
\newcommand{\la}{\langle}
\makeindex
\begin{document}

\section{Introduction}

Quantum distribution functions were first introduced by Wigner \cite{Wigner:1932eb}
 as functions on classical phase space encoding, for the mean values of quantum observables,
the same information as a quantum state. More precisely, for a given quantum state $|\psi\ra$, the corresponding quantum distribution function $F_{|\psi\ra}(p,q)$ is defined by requiring
\be
\la \psi |\hat{A}|\psi \ra= \int dp dq F_{|\psi\ra}(p,q) A(p,q)
\ee
where $A(p,q)$ is the phase space function corresponding to the operator $\hat{A}$. Different quantum distribution functions correspond to different maps $A(p,q)\to \hat{A}$. The most popular is the Wigner distribution function, corresponding to the Weyl map \cite{Wey31}.

After the discovery of the holomorphic anomaly \cite{Bershadsky:1993ta,Bershadsky:1993cx} for topological strings it was suggested in \cite{Witten:1993ed} to relate the topological string amplitudes with quantum wave functions of an auxiliary quantum mechanical system with classical phase space $H^3(M,\real)$, where $M$ is the Calabi-Yau target space of B model topological strings. This suggestion in \cite{Witten:1993ed} was introduced as a attempt to deal with the problem of quantum background independence in string theory.

Recently in \cite{Ooguri:2004zv} (OSV) it was suggested the fascinating conjecture that black hole entropy for BPS Reissner-Nordstrom type black holes in $\Ene=2$ supergravity is directly related with the quantum Wigner distribution function $F^W_{|\psi_{\rm top}\ra}(p,q)$ defined on $H^3(M,\real)$ for the state $|\psi_{\rm top}\ra$ associated with the topological string amplitudes with target space $M$. This conjecture is important for two reasons. First, for pointing out a deep connection between topological strings and black holes. Secondly because $F^W_{|\psi_{\rm top}\ra}(p,q)$ leads to unexpected corrections to the black hole entropy beyond those encoded in the Bekenstein-Hawking-Wald (BHW) entropy formula \cite{Wald:1993nt,Iyer:1994ys,Jacobson:1993vj,LopesCardoso:1998wt,LopesCardoso:1999cv,LopesCardoso:1999ur,LopesCardoso:1999xn}. 

The physical interpretation of quantum distribution functions is far from clear and, moreover, for the same quantum state $|\psi\ra$, one can define different quantum distribution functions $F_{|\psi\ra}$ on classical phase space. Why black hole entropy is related to Wigner distribution function and what is the physical meaning of OSV corrections to the black hole entropy are in this sense intimately related questions.

The OSV conjecture is based on the definition of a mixed black hole ensemble with fixed charge $p$ and potential $\phi$
\be\label{oosv}
Z_{BH}(p,\phi) = \sum_{q} \Omega(p,q)e^{q \phi}
\een
where $\Omega(p,q)$ is counting the black hole microscopic degeneracies. The OSV conjecture establishes a relation between this mixed black hole partition function and the topological string partition function at the attractor geometry 
\be\label{osv}
Z_{BH}(p,\phi) = |Z_{top}|^{2}
\een
This conjecture implies that the macroscopic Bekenstein-Hawking-Wald entropy is the Legendre transform of the black hole free energy $\log Z_{BH}(p,\phi)$. In fact the conjecture was motivated by interpreting the macroscopic supergravity  formula for the BHW entropy as the typical relation, in the thermodynamic limit (to be defined here as the large black hole charge limit), between different statistical ensembles. Recently different attempts for a direct derivation of (\ref{osv}) using a M-theory lift has been suggested in \cite{Gaiotto:2006ns,Beasley:2006us}. By inverting (\ref{oosv}) and  by using a quantum mechanical wave function interpretation of $Z_{top}$, an integral representation of the microscopic degeneracies 
$\Omega(p,q)$ can be derived. This integral representation is formally similar to a Wigner distribution function on $H^{3}(M,\real)$.

In this note we will address the OSV conjecture from a different point of view, namely we will
start with the BHW macroscopic entropy and we will identify this entropy with a quantum distribution function on $H^{3}(M,\real)$ based on coherent states. Later on we will use an integral representation of this distribution function with kernel the Wigner function and we will study the different regimes in which this integral representation becomes equivalent to OSV.

More precisely, by using the Kähler quantization
on $H^{3}(M)$, we define a map from from black hole data (charges and attractor geometry) into coherent states. This map leads to an explicit representation of the BHW entropy as a mixed Husimi/anti-Husimi distribution. The mixed nature of this quantum distribution comes from the different normal ordering prescriptions for the creation and annihilation operators associated with the Kähler coordinates. Moreover the
Husimi distribution function can be related to the Wigner distribution by
\be\label{husimi}
F^H(p,q) = \sum_{p,q} F^{W}(p^\prime,q^\prime) g(p-p^\prime,q-q^\prime)
\ee
with $g$ a gaussian function with width the one of the coherent state entering into the definition of $F^H$.
This second equation is the analogous of equation (\ref{oosv}) and will be used to provide a precise relation, in certain limits, between the OSV entropy $\Omega(p,q)$ and the Wigner distribution function.
The physical meaning of (\ref{husimi}) is that of a gaussian smearing or coarse graining of Wigner distributions. The same type of formal coarse graining relation appears between the BHW and the OSV entropies in the limit of attractor complex structure moduli at infinity.

The outline of the paper is as follows. In section 2 we carefully review the connection between topological string amplitudes and the quantum mechanics defined on $H^3(M,\real)$ using Kähler and real polarizations. In section 3 we review the BHW entropy formula and the OSV generalization paying special attention to non-holomorphic contributions. In section 4 we define a map from BPS RN type black holes into coherent states and we represent the BHW entropy in terms of a mixed Husimi/anti-Husimi quantum distribution function. We describe the connection with Wigner distribution and we identify the limits where we can formally recover OSV. We conclude with a short discussion.

\section{Topological Strings and the Quantization of {\boldmath$H^3(CY_3,\real)$}}

\subsection{{\boldmath$H^3(CY_3,\real)$} as a Phase Space}

In order to review the precise map between topological strings and the geometric quantization of $H^3(M,\real)$ \cite{Witten:1993ed,Gerasimov:2004yx,Verlinde:2004ck,Loran:2005ma}, let us consider a 7d field theory action on $M \times \real$, being $M$ a compact Calabi-Yau threefold, with action functional
\be
\label{act}
S(C)=2 \int_{M \times \real} C \wedge d^\prime C 
\ee
where $C$ is a real 3-form on $M \times \real$ and $d^\prime$ is the 7d exterior derivative. This action can be seen as a map $S:\Efe \to \real$ from the space $\Efe$ of kinematically allowed field configurations to $\real$ with lagrangian $\Ele= 2 C \wedge d^\prime C $ as a 7-form. Under a variation of $C$ 
\be
\delta \Ele = E \wedge \delta C + d^\prime \Theta [\delta C]
\ee
where
\be
E=+4 d^\prime C =0
\ee
are the classical equation of motion, and
\be
\Theta [\delta C] = -2 C \wedge \delta C
\ee
is the symplectic potential current density. From a 6d point of view we can decompose
\be
C= \gamma + \omega \wedge d t^\prime
\ee
where $\omega$ and $\gamma$ are real 2-forms and 3-forms on $M$. Denoting the 6d exterior derivative as $d$, and the derivates with respect to $t^\prime$ as $\cdot$, we can write the action in terms of a 6-form lagrangian on $M$
\be
S=\int_{M\times\real} \Ele_{6d} \wedge dt\prime 
\ee
\be
\Ele_{6d}=2\gamma (-\dot{\gamma}+d\omega) + 2 \omega \wedge d\gamma
\ee
The classical equations of motion are
\bea
E_\gamma&=&4\left( \dot{\gamma}-d\omega  \right)=0\\
E_\omega&=&4d\gamma=0
\eea
and we have invariance with respect to the transformations
\bea
\gamma &\to& \gamma + d \beta\\
\omega &\to& \omega + \dot{\beta} + d\epsilon
\eea

In order to obtain the physical phase space of the theory we follow formally \cite{Lee:1990nz}. If we consider a 2-parameter family of variations $\delta_1, \delta_2$
\be
\delta_1 \delta_2 \Ele = \delta_1 E \wedge \delta_2 C + E \wedge \delta_1\delta_2 C + \delta_1d^\prime \Theta_2
\ee
the symplectic current 6-form is defined by
\be
\omega(\delta_1C,\delta_2C)\equiv \delta_1 \Theta_2 - \delta_2 \Theta_1 = -4 \delta_1 C \wedge \delta_2 C
\ee
implying
\be
d^\prime  \omega(\delta_1C,\delta_2C) = \delta_2 E \wedge \delta_1 C - \delta_1 E \wedge \delta_2 C
\ee
thus $\omega(\delta_1C,\delta_2C)$ is conserved if the variations $\delta_1 C$ and $\delta_2 C$ parametrize a 2-parameter family of solutions. In this case, the functional
\be
\label{sympform}
\Omega(1,2) = \int_M \omega(\delta_1C,\delta_2C) = -4 \int_M \delta_1 \gamma \wedge \delta_2 \gamma
\ee
which defines a 2-form on the space of field configurations of the theory, is independent on $t^\prime$. The physical phase space is the subspace on which $\Omega(1,2)$ is non-degenerate. Since the conjugate momenta are
\bea
\pi_\gamma&=& \frac{\pd\Ele_{6d}}{\pd \dot{\gamma}}=-2\gamma\\
\pi_\omega&=& \frac{\pd\Ele_{6d}}{\pd \dot{\omega}}=0
\eea
one needs, in order to identify the physical phase space, to work with the hamiltonian
\be
\Hache_{6d}(\gamma,\omega,\pi_\gamma,\pi_\omega)=
-4\gamma\wedge d\omega - d(\gamma\wedge\omega)
\ee
and the constraints
\bea
\Phi^{(1)}_\gamma&=&\pi_\gamma + 2\gamma\\
\Phi^{(1)}_\omega&=&\pi_\omega\\
\Phi^{(2)}_\omega&=&4d\gamma
\eea
$\Phi^{(1)}_\gamma$ and $\Phi^{(1)}_\omega$ are primary constraints, whereas $\Phi^{(2)}_\omega$ is a secondary constraint obtained from $\dot{\Phi}^{(1)}_\omega=0$. Both $\Phi^{(1)}_\omega$ and $\Phi^{(2)}_\omega$ are first class constraints and one has to take into account this fact in order to quantize the theory. On the other hand, from $\dot{\Phi}^{(1)}_\gamma=0$ one obtains $v_\gamma=d\omega$ and, therefore, $\Phi^{(1)}_\gamma$ is a set of second class constraints, implying that one has to work with Dirac brackets instead that with Poisson brackets.

After imposing these constraints one can see that $\Omega(1,2)$ is non-degenerate if we restrict the field configurations to be elements  $\gamma \in H^3(M,\real)$ and, hence, $\Omega(1,2)$ is a symplectic form giving a phase space structure to $H^3(M,\real)$.
In order to quantize the theory one can follow the  geometric quantization procedure \cite{Witten:1993ed}. This amounts to choose a complex structure on $M$ that induces a polarization on $H^3(M,\real)$. For the  7d field theory (\ref{act}), this procedure is equivalent to work with the Kähler coordinates $(\l^{-1},x^i)$
\be
\label{kaco}
\g= \frac{1}{2} \left[ \l^{-1}\Omega + x^i \De_i \Omega + cc   \right]
\ee
where $\Omega$ is the holomorphic 3-form on $M$ in the chosen complex structure (and in a chosen section of the line bundle $\Ele$), with
\be
\De_i \Omega = \left( \frac{\pd}{\pd t^i} + \pd_i K \right) \Omega
\ee
the covariant derivative for sections of  $\Ele$ and $t^i$ the coordinates on $\Eme$, whereas
\be
X^I=\int_{A_I}\Omega
\ee
are projective coordinates on $\Eme$, i.e  sections of $\Ele$. The Kähler potential of $\Eme$ is
\be
K=-\log \left[ i\left( \bar{X}^IF_I - X^I\bar{F}_I  \right)  \right]
\ee
where
\bea
F_I&=&\int_{B^I} \Omega = \frac{\pd F_0}{\pd X^I}\\
\t_{IJ}&=&\frac{\pd F_J}{\pd X^I}
\eea
being $F_0(X)$ the prepotential of $M$. In the coordinates (\ref{kaco}) the constraint
$\Phi^{(1)}_\gamma=\pi_\gamma + 2\gamma$ becomes
\bea
\pi_{\l^{-1}}&\equiv &\pi=-\frac{i}{2}e^{-K}\bar{\l}^{-1}\\
\pi_{\bar{\l}^{-1}}&\equiv &\bar{\pi}=+\frac{i}{2}e^{-K}{\l}^{-1}\\
\pi_{x^i}&\equiv &\pi_i=+\frac{i}{2}e^{-K}G_{i\bar{j}}\bar{x}^{\bar{j}}\\
\pi_{\bar{x}^{\bar{i}}}&\equiv &\bar{\pi}_{\bar{i}}=-\frac{i}{2}e^{-K}G_{j\bar{i}}{x}^{{j}}
\eea
The corresponding Dirac brackets are \cite{Loran:2005ma}
\bea
\{ \l^{-1},\pi \}_D&=&\frac{1}{2}\\
\{ \bar{\l}^{-1},\bar{\pi} \}_D&=&\frac{1}{2}\\
\{ x^i,\pi_j \}_D&=&\frac{1}{2}\d^i_j\\
\{ \bar{x}^{\bar{i}},\bar{\pi}_{\bar{j}} \}_D&=&\frac{1}{2}\d^{\bar{i}}_{\bar{j}}
\eea
leading to the quantum commutators
\bea
\left[\l^{-1},\bar{\l}^{-1}\right]&=&-\hbar e^K\\
\left[x^i,\bar{x}^{\bar{j}}\right]&=&\hbar e^K G^{i\bar{j}}
\eea

\subsection{Coherent States}

Notice that, since the metric of $\Eme$ is definite-positive, $\bar{\l}^{-1}$ and $x^i$ act as annihilation operators. Defining \cite{Verlinde:2004ck} $Q\equiv 2i\pi=e^{-K}\bar{\l}^{-1}$, we can work with the states $| x,Q \ra$:
\bea
\hat{Q}| x,Q \ra&=&Q| x,Q \ra\\
\hat{x}^i | x,Q \ra&=&x^i | x,Q \ra
\eea
However, in order to establish the connection with topological strings, it is useful to work formally with the non-normalizable states $| x,\l^{-1} \ra$ defined by:
\bea
\hat{\l}^{-1}| x,\l^{-1} \ra&=&\l^{-1}| x,\l^{-1} \ra\\
\hat{x}^i | x,\l^{-1} \ra&=&x^i | x,\l^{-1} \ra
\eea
As it is standard in quantum mechanics we can define the coherent states in terms of the
``vacuum'', $| 0,0\ra$ in our case. This allows us to define a relative normalization with respect to 
$| 0,0\ra$. If we formally define the coherent state $|\alpha\ra$ as $e^{\alpha a^{+}}|0 \ra$ we will
get
\bea
\label{badco}
| x,\l^{-1} \ra= \exp \left[ -\frac{1}{\hbar} e^{-K} \hat{\bar{\lambda}}^{-1}{\lambda}^{-1} + \frac{1}{\hbar} e^{-K} x^i\hat{\bar{x}}^{\bar{j}} G_{i\bar{j}}   \right] | 0,0 \ra\\
\imaginario=\int d\mu_{x,\l^{-1}} \exp \left[ +\frac{1}{\hbar} e^{-K} \bar{\lambda}^{-1}{\lambda}^{-1} - \frac{1}{\hbar} e^{-K} x^i{\bar{x}}^{\bar{j}} G_{i\bar{j}}   \right] | x,\l^{-1} \ra \la \bar{x},\bar{\l}^{-1} |\\
\frac{\la \bar{x}^\prime,\bar{\l}^{-1\prime} | x,\l^{-1} \ra }{\la \bar{0},\bar{0} |0,0 \ra }=\exp \left[ -\frac{1}{\hbar} e^{-K} {\bar{\lambda}}^{-1\prime}{\lambda}^{-1} + \frac{1}{\hbar} e^{-K} x^i{\bar{x}}^{\bar{j}\prime} G_{i\bar{j}}   \right]
\eea
where $d\mu_{x,\l^{-1}}\propto d^{h}xd^h\bar{x} d\l^{-1}d\bar{\l}^{-1}$, with a factor of proportionality we do not fix for the moment, due to the fact that $| 0,0 \ra$ is non-normalizable. The crucial point about
this way to define the coherent states is that it ensures a holomorphic dependence with
respect to $(x, \lambda^{-1})$. The price to be paid is that this definition leads to a ``bad'' relative normalization, namely
\be
\frac{\la \bar{x},\bar{\l}^{-1} | x,\l^{-1} \ra }{\la \bar{0},\bar{0} |0,0 \ra }=\exp \left[ -\frac{1}{\hbar} e^{-K} {\bar{\lambda}}^{-1}{\lambda}^{-1} + \frac{1}{\hbar} e^{-K} x^i{\bar{x}}^{\bar{j}} G_{i\bar{j}}   \right]
\ee
is clearly different to one. 

Another way of describing these states is by using big phase space variables
$x^I=\l^{-1}X^I + x^i \De_i X^I$. That is,
\be
\label{igua}
| {x}^I \ra= |x^i,\l^{-1}\ra
\ee
Notice that one has to choose a particular symplectic homology basis in order to work with big phase space variables. The quantization rule in these variables is
\be
\left[ x^I,\bar{x}^J \right] = \hbar \left[ \I \t(X)  \right]^{-1IJ}
\ee
Notice however that, in order to get (\ref{igua}), one must define $| {x}^I \ra$ using a different Planck constant, concretely $2\hbar$.

One can also work with real polarization coordinates
\be
\g=p^I\a_I + q_I \bet^I \in H^3(M,\real)
\ee
Since
\bea
p^I &=& \R x^I \\
q_I &=& \R \left[ \t_{IJ}(X) x^J  \right] 
\eea
we can understand this representation as the $\I \tau \to \infty$ limit of the big phase space representation. Although the commutators are
\be
\left[ q_I , p^J  \right]=  i \frac{\hbar}{2} \d^J_I
\ee
we will define the states $|p\ra$ using the modified Planck constant $2\hbar$. Notice that, from the point of view of this real representation, the states $|x\ra$ are in fact coherent states centered around the point $(p,q)$ in phase space and with a width, measuring the quantum resolution,  given by the base holomorphic 3-form $\Omega$ used to define the Kähler polarization
\bea
(\Delta q)_I(\Delta q)_J&=&\frac{\hbar}{2} \left[ \left(\imaginario + \R\t(X)(\I \t(X))^{-1}    \right)   \I \t(X)\left( \imaginario + (\I \t(X))^{-1}\R\t(X)  \right)  \right]\nonumber\\
(\Delta p)^I(\Delta p)^J&=&\frac{\hbar}{2}(\I \t(X))^{-1} \label{wi}
\eea
Moreover  the quantities $\R \t(X)$ play the role of squeezed parameters, leading to a quantum uncertainty $\Delta q_I \Delta p^I$ (without summing)  bigger than $\hbar/2$.

\subsection{Topological Strings Amplitudes}

Since the widths (\ref{wi}) of the coherent states depend on the base complex structure $\Omega$, these states will transform under variations of $\Omega$. These transformations \cite{Witten:1993ed,Verlinde:2004ck}  can be interpreted as a Bogolioubov transformation and are exactly equal to the holomorphic anomaly equations \cite{Bershadsky:1993cx} governing the background dependence of the generating function of B model topological string correlators on $M$. More precisely, if one defines the topological string function \cite{Verlinde:2004ck} as
\be
\psi_{\mbox{\scriptsize{top}}}(\l^{-1},x^i;t,\bar{t})=e^{F_1(t,\bar{t})-\bar{f}_1(\bar{t})}\psi_{\rm gen}(\l,\l x; t,\bar{t})
\ee
where $\bar{f}_1(\bar{t})$ is the antiholomorphic part of the genus one free energy $F_1$, and where
\be
\psi_{\rm gen}(\l,x; t,\bar{t})=\l^{\chi/24-1}\exp \left[\sum_{g=0}\l^{2g-2} \sum_{n=0} \frac{1}{n!} C^g_{i_1,...,i_n}(t,\bar{t})x^{i_1}...x^{i_n} \right]
\ee
is the generating function of correlators, one finds that $\psi_{\mbox{\scriptsize{top}}}$ varies with respect to $(t,\bar{t})$ in the same way as $|x^i,\l^{-1}\ra_{t,\bar{t}}$. It is important here to stress that
in order to match the transformation of the coherent state with the holomorphic anomaly equation it
is crucial to use the coherent states defined in (\ref{badco}) i.e those that depend holomorphically on
$(x, \lambda^{-1})$.  This suggest to define an state 
$|  \psi_{\rm{top}} \ra$ such that
\be
\psi_{\mbox{\scriptsize{top}}}(x,\l^{-1};t,\bar{t})=\la \psi_{\mbox{\scriptsize{top}}} |x,\l^{-1}\ra_{t,\bar{t}}
\ee
that is, $\la \psi_{\mbox{\scriptsize{top}}} |$ is a quantum state in the Hilbert space defining the quantization of $H^3(M,\real)$ that contains all the information about topological string correlators. Moreover it can be shown \cite{Gerasimov:2004yx} that the requirement on $\la \psi_{\mbox{\scriptsize{top}}} |$ to be a physical state, that is, one that satisfies
\be
\label{confis}
\hat{\Phi}^{(2)}_\omega| \psi_{{\rm phys}} \ra \equiv \widehat{d\gamma} | \psi_{{\rm phys}} \ra=0
\ee
leads to a direct connection between $\la \psi_{\mbox{\scriptsize{top}}} |x,\l^{-1}\ra_{t,\bar{t}}$ and the topological string generating function.
The sketch of the proof is as follows. Before restricting to $H^3(M,\real)$, one decomposes the operator $\gamma$ into
\be
\gamma = \gamma^{3,0}\oplus\gamma^{2,1}\oplus\gamma^{1,2}\oplus\gamma^{0,3}
\ee
by choosing a concrete complex structure (whose coordinates in $\Eme$ are $(t,\bar{t})$). The symplectic form (\ref{sympform}) implies that $\gamma^{2,1}$ does not commute with $\gamma^{1,2}$, and the same for $\gamma^{3,0}$ with $\gamma^{0,3}$. We can now write the condition (\ref{confis})
\bea
\left(\hat{\pd \gamma}^{2,1} + \hat{\bar{\pd} \gamma}^{3,0}\right)| \psi_{{\rm phys}} \ra=0\\
\left(\hat{\pd \gamma}^{1,2} + \hat{\bar{\pd} \gamma}^{2,1}\right)| \psi_{{\rm phys}} \ra=0\\
\left(\hat{\pd \gamma}^{0,3} + \hat{\bar{\pd} \gamma}^{1,2}\right)| \psi_{{\rm phys}} \ra=0\\
\eea
They imply that the space of physical states can be obtained from the naive Hilbert space by applying the hermitian projector
\be
|\psi_{{\rm phys}}\ra= \hat{\Pi}|\psi_{0}\ra
\ee
where
\be
\hat{\Pi}=\int D\Lambda D\sigma D b \exp \int_{M}\left[ \Lambda \wedge \left(\hat{\pd \gamma}^{2,1} + \hat{\bar{\pd} \gamma}^{3,0}\right) + \sigma \wedge \left(\hat{\pd \gamma}^{0,3} + \hat{\bar{\pd} \gamma}^{1,2}\right) + b \wedge \left(\hat{\pd \gamma}^{1,2} + \hat{\bar{\pd} \gamma}^{2,1}\right) \right]
\ee
Now, we apply this projection operator to the coherent states $|\gamma^{3,0},\gamma^{2,1}\ra$. With the ``bad'' normalization (\ref{badco}), we have
\bea
\hat{ \gamma}^{0,3} | \gamma^{3,0},\gamma^{2,1}\ra=C_{3,0} \frac{\d}{\d \gamma^{3,0}}| \gamma^{3,0},\gamma^{2,1}\ra\\
\hat{ \gamma}^{1,2} | \gamma^{3,0},\gamma^{2,1}\ra=C_{2,1} \frac{\d}{\d \gamma^{2,1}}| \gamma^{3,0},\gamma^{2,1}\ra\\
\eea
where $C_{3,0}$ and $C_{2,1}$ are the commutators of $\gamma^{3,0}$ and $\gamma^{2,1}$ with their hermitian conjugate operators, we obtain
\bea
\hat{\Pi} | \gamma^{3,0},\gamma^{2,1}\ra&=& \int D\Lambda D\sigma D b \exp \int_{M}\left[ \Lambda \wedge \left(\hat{\pd \gamma}^{2,1}+ \hat{\bar{\pd} \gamma}^{3,0}\right)+\bar{\pd} b \wedge \gamma^{2,1}   -\frac{C_{2,1}}{2} \bar{\pd}b\pd b \right]\nonumber\\
&& | \gamma^{3,0}+C_{3,0} \pd \sigma,\gamma^{2,1} + C_{2,1} \pd b - C_{2,1} \bar{\pd}\sigma\ra
\eea
If we also choose a section in the line bundle $\Ele$, we can parametrize $\gamma^{3,0}$ and $\gamma^{2,1}$ in Kähler coordinates
\bea
\gamma^{3,0}&=&\frac{1}{2}\l^{-1}\Omega + \pd \chi\\
\gamma^{2,1}&=&\frac{1}{2}x^i\De_i \Omega + \pd \xi + \pd^\dagger \tilde{\xi}-\bar{\pd} \chi
\eea
In this coordinates
\be
\hat{\Pi} | \gamma^{3,0},\gamma^{2,1}\ra= \d (\pd\pd^\dagger \tilde{\xi})\int D\sigma D b \exp \int_{M}\left[ \bar{\pd} b \wedge \pd \xi   -\frac{C_{2,1}}{2} \bar{\pd}b\pd b \right] | \gamma^{3,0}+C_{3,0} \pd \sigma,\gamma^{2,1} + C_{2,1} \pd b - C_{2,1} \bar{\pd}\sigma\ra
\ee
Restricting the dependence only in the physical coordinates $\l^{-1},x^i$
\be
\hat{\Pi} | \l^{-1},x^i\ra= \int D\sigma D b \exp \int_{M}\left[ -\frac{C_{2,1}}{2} \bar{\pd}b\pd b \right] | \l^{-1}+C_{3,0} \l^{-1}_{\pd \sigma},x^i + C_{2,1} x^i_{\pd b} - C_{2,1} x^i_{\bar{\pd}\sigma}\ra
\ee
where
\bea
\pd \sigma&=&\frac{1}{2}\l^{-1}_{\pd\sigma}\Omega \\
\bar{\pd}\sigma&=&\frac{1}{2}x^i_{\bar{\pd} \sigma}\De_i \Omega \\
\pd b&=&\frac{1}{2}x^i_{\pd b}\De_i \Omega 
\eea
The only restriction we impose to $|\psi_{0}\ra$ is that $\la \psi_{0}|\hat{\Pi}| \l^{-1},x\ra$ converges.
By choosing
\be
\label{psicero}
\la \psi_{0}| \l^{-1},x\ra=\exp \left[ f_1(t) + \frac{1}{6\l^{-1}} C_{ijk}^{0}(t)x^ix^jx^k  \right]  
\ee
one obtains that the corresponding physical state is $\exp f_1$ times the Kodaira-Spencer partition function. But the Kodaira-Spencer theory \cite{Bershadsky:1993cx} is the string field theory of the B model, in such a way that its partition function gives the functional generator of correlation functions. Therefore
\be
\la \psi_{\mbox{\scriptsize{phys}}} |x,\l^{-1}\ra_{t,\bar{t}} = 
\la \psi_{0} | \hat{\Pi} |x,\l^{-1}\ra_{t,\bar{t}}=
\psi_{\mbox{\scriptsize{top}}}(x,\l^{-1};t,\bar{t})
\ee

\section{Bekenstein-Hawking-Wald and Ooguri-Strominger-Vafa Entropies}

Let us consider 4d BPS dyonic black holes in type IIB compactifications on $M$. From the microscopic point of view, they correspond to D3-branes wrapping a CY homology 3 cycle
\be
\Ce_{p,q}=q_IA_I - p^JB^J \in H_3(M,\entero)
\ee
where $(A_I,B^J)$ is a sympletic basis of $H^3(M,\entero)$. From the macroscopic point of view, they are BPS Reissner-Nordstrom type solutions of
the 4d $\Ene=2$ effective supergravity theory whose vector multiplet sector contains $n_v=h_{2,1}$ abelian vector multiplets, each one containing 1 vector gauge field, 1 $SU(2)$ doublet of chiral fermions and 1 complex scalar $z^i$, parameterizing the moduli space $\Eme$ of complex structures on $M$. These black holes have $p^I$ and $q_I$ magnetic and electric charges under the gauge fields obtained from the reduction of $F_5$ over 3-cycles on $M$.

It is known that these BH solutions present the so called attractor phenomenon \cite{Ferrara:1995ih,Strominger:1996kf,Ferrara:1997tw}. Each point in $\Eme$ fixes a particular Hodge decomposition
\be
H^3(M)=H^{3,0}\oplus H^{2,1}\oplus H^{1,2}\oplus H^{0,3}
\ee
The attractor mechanism fixes the vector multiplet moduli at the horizon in terms of the charges $p,q$ in such a way that \cite{Moore:2004fg}
\be 
\label{claatt}
\gamma_{p,q} = \R \left( \l^{-1} \Omega \right) \in H^{3,0} \oplus H^{0,3}
\ee
for the corresponding Hodge decomposition, where $\gamma_{p,q}$ is the Poincaré dual 3-form of $\Ce_{p,q}$. Condition (\ref{claatt}) is equivalent to the attractor equations 
\bea
p^I&=& \R \left( X^I \l^{-1}\right)\\
q_I&=& \R \left( F_I \l^{-1} \right)
\eea
$\l$ is a section of $\Ele$ inserted in order to have a Kähler gauge invariant expression.

A consequence of the attractor mechanism is that the Bekenstein-Hawking entropy does not depend on the asymptotic values of the vector multiplet moduli. Defining $\phi=2\I X/\l $ the Bekenstein-Hawking entropy is given by
\bea
S_{\rm BH}&=&\left. -\frac{\pi}{2} \I \t_{IJ} (X) \frac{X^I}{\l}\frac{\bar{X}^I}{\l} \right|_{p,q}=\nonumber\\
&=&i\frac{\pi}{2}F_0\left( \frac{X}{\l}  \right)-i\frac{\pi}{2}\bar{F}_0\left( \frac{\bar{X}}{\bar{\l}}  \right)+\frac{\pi}{2}q_I \phi^I
\label{BHentro}
\eea
where (\ref{BHentro}) is evaluated at the attractor value $X_{p,q}^I\l^{-1}_{p,q}=p^I+i\phi^I_{p,q}/2$ solving (\ref{claatt}).

In the case we are considering, type IIB compactifications on $M$, we can also obtain, from string theory, additional genus g corrections to the 4d effective supergravity theory. The physical amplitudes which give rise to some of these corrections are given by B-model topological strings \cite{Bershadsky:1993cx,Antoniadis:1993ze}. 
With these corrections the vector multiplet part of the 4d effective action is fully specified by the quantum corrected supergravity prepotential, which is given in terms of the B-model topological string free energy by
\be\label{prep}
F_{{\rm mod}} (\tilde{X}\l^{-1},\bar{\tilde{X}}\bar{\l}^{-1},\Gamma=256,\bar{\Gamma}=256)=-\frac{2}{\pi}i \sum_{g} \left(\frac{\lambda}{X^0}\right)^{2g-2}F_{\rm top}^{g} (t,\bar{t})
\ee
where $\tilde{X}$ are the $h^{2,1}+1$ chiral superfields constructed from the vector multiplets and $\tilde{\Gamma}$ is the chiral superfield obtained from the Weyl multiplet whose lowest component is the squared of the graviphoton field strength.
Notice that the modified quantum corrected prepotential defined above is obtained from the genus expansion of the holomorphic one by replacing the different genus contributions by the corresponding solutions to the holomorphic anomaly equations\footnote{In formula (\ref{prep}) $t,\bar{t}$ are non-projective coordinates of the moduli space point given by the projective ones $X,\bar{X}$. Because of $\lambda$, both sides of the equality are Kähler gauge invariant. Therefore we can work in every Kähler gauge, for instance, $X^0=1$}. The holomorphic anomaly, fixing the different genus contributions $F_{\rm top}^{g}(t, \bar t)$ to the topological string amplitude, is directly related to the symplectic anomaly for the string loop corrected supergravity prepotential \cite{deWit:1996wq,LopesCardoso:1999ur,LopesCardoso:2004xf,LopesCardoso:2006bg,Aganagic:2006wq}. The quantum field theory understanding of this non holomorphicity comes from the difference between the 1PI effective action and the Wilsonian action. 

Using this modified prepotential and Wald's expression \cite{Wald:1993nt,Iyer:1994ys}  for the entropy one obtains
\be
\label{nhBHW}
S_{\rm BHW}(p,q)=
 F_{{\rm top}}(\l,t,\bar{t}) +\bar{F}_{\rm top}(\bar{\l},\bar{t},t)+\frac{\pi}{2}q_I\phi^I 
\ee
where now this expression is evaluated at the corrected attractor values, to be denoted  $X_{p,q}^{\rm quan}$ and $\l_{p,q}^{\rm quan}$, solving the following
generalized (non-holomorphic) attractor equations\footnote{See \cite{Mohaupt:2000mj,LopesCardoso:2006bg} and references therein}
\bea
p^I&=& \R \left( X^I\l^{-1} \right)\\
q_I&=& \R \left( \frac{\pd F_{{\rm mod}} (X\l^{-1},\bar{X}\bar{\l}^{-1},\Gamma,\bar{\Gamma})}{\pd (X^I\l^{-1})} -\frac{\pd F_{{\rm mod}} (X\l^{-1},\bar{X}\bar{\l}^{-1},\Gamma,\bar{\Gamma})}{\pd (\bar{X}^I\bar{\l}^{-1})}\right)\\
\Gamma & = & 256
\eea
Notice the extra term $-\frac{\pd F_{{\rm mod}} }{\pd (\bar{X}^I\bar{\l}^{-1})}$ that appears in the second attractor equation coming from the non-holomorphic corrections to the prepotential\footnote{An important point to discuss is that these attractor equations are not the same as the one that appears in the literature. We are using the formulas (2.12) and (3.4) of \cite{LopesCardoso:2006bg}, but we are calling $F_{\rm mod}$ to the quantity $F+i\Omega$ of \cite{LopesCardoso:2006bg} instead of $F+2i\Omega$.}. In what follows we are going to consider that the genus 1 free energy that appears into the BHW formula does not contain the term $\bar{f}_1(\bar{t})$ (see \cite{Sen:2004dp,Sen:2005pu}).

As it was first noticed in \cite{Ooguri:2004zv}, equation (\ref{nhBHW}) identifies the BHW entropy with the Legendre transform of $F_{{\rm top}}(\l,t,\bar{t}) +\bar{F}_{\rm top}(\bar{\l},\bar{t},t)$. This can be interpreted as the typical
relation, in the thermodynamical limit, between the microcanonical and the canonical ensemble, where here the ``canonical'' ensemble variable is the electric potential $\phi$ and the ``thermodynamical limit'' is the limit of large BH charges. This formal analogy leads to the OSV conjecture that the partition function of
the mixed ensemble with $(p,\phi)$ fixed 
\be
Z_{BH}(p,\phi) = \sum_{q}\Omega(p,q) e^{\frac{\pi}{2}q \phi}
\ee
where $\Omega(p,q)$ counts the number of microscopic BPS states, would be given  in terms of the topological string amplitude by
\be
\label{OSVcon}
Z_{\rm BH}(p,\phi)=\left| \exp F_{{\rm top}}(\l,t,\bar{t})   \right|^2 
\ee
This conjecture has been studied and refined in 
\cite{Vafa:2004qa,Dabholkar:2004yr,Aganagic:2004js,Dabholkar:2005by,Ooguri:2005vr,Aganagic:2005dh,Dijkgraaf:2005bp,Sen:2005wa,Dabholkar:2005dt,Shih:2005he,Aganagic:2005wn,Gunaydin:2005mx,Jafferis:2006ny,Gaiotto:2006wm,deBoer:2006vg,Verlinde:2004ck,Gaiotto:2006ns,Beasley:2006us,LopesCardoso:2004xf,LopesCardoso:2006bg}. 
From this OSV conjecture we can derive the microcanonical microscopic entropy in terms of the topological string amplitude, namely
\be
\label{SOSV}
\Omega(p,q)=\exp S_{\rm OSV}(p,q)=C \int_{C_\Eme} d\phi \left| \exp F_{{\rm top}}(\l,t,\bar{t})   \right|^2 e^{\frac{\pi}{2}q\phi}
\ee
with $X\l^{-1}=p+i\frac{\phi}{2}$ and where $C_\Eme$ is the path in $\Eme$ (concretely of its line bundle) corresponding to points at which $\R (X\l^{-1})=p$ and where $C$ is some function of the charges ensuring the whole formula is symplectic invariant (see for example \cite{LopesCardoso:2006bg}).
In what follows we will concentrate our attention in the quantum mechanical meaning of equation (\ref{OSVcon}) as defining a black hole quantum distribution function on $H^3(M)$. Once this is done we will try to unravel the physical meaning of (\ref{SOSV}) using the relation between the black hole quantum distribution function on $H^3(M)$ and the Wigner distribution function.

\section{Black Hole Quantum Distribution Function}

\subsection{Bekenstein-Hawking-Wald Entropy and the Quantization of $H^3$}

The first step in our construction is to define a map that associate to a given BPS BH of charges (p,q) a concrete coherent state $| BH(p,q) \ra$ in the Kähler quantization of $H^3(M,\real)$. Intuitively $| BH(p,q) \ra$ is just a coherent state centered in phase space on the point defined by the BH charges and with a quantum resolution determined by the complex structure at the attractor point. More precisely, denoting $X^{\rm quan}_{p,q}/\l_{p,q}^{\rm quan}$ the solutions of the corrected attractor equations for a BH with charges $(p,q)$, we define a map $(p,q)\to(\tilde{p},\tilde{q})$ such that
\be
\frac{X^{}_{\tilde{p},\tilde{q}}}{\l_{\tilde{p},\tilde{q}}^{}}=\frac{X^{\rm quan}_{p,q}}{\l_{p,q}^{\rm quan}}
\ee
where $X_{\tilde{p},\tilde{q}}$ and $\lambda_{\tilde{p},\tilde{q}}$ are the solutions to the corresponding classical attractor equations for charges $(\tilde{p}, \tilde{q})$. 
Obviously, $\tilde{p}=p$, but $\tilde{q}=\tilde{q}(p,q)$ will be in general different from $q$, reflecting the quantum dressing of the BH coherent state. Notice that, if there is solution to the corrected attractor equations, one can always define $\tilde{q}$ for a given $p,q$. The coherent state we associate with the BH of charges $(p,q)$ is the one centered in $(p,\tilde{q})$ with base point given by $X^{}_{{p},\tilde{q}}$:
\be
BH(p,q) \longrightarrow | x_{p,\tilde{q},X^{}_{{p},\tilde{q}}},\l^{-1}_{p,\tilde{q},X^{}_{{p},\tilde{q}}} \ra_{X^{}_{{p},\tilde{q}},\bar{X}^{}_{{p},\tilde{q}}} 
\ee
Notice that $x_{p,\tilde{q},X^{}_{{p},\tilde{q}}}=0$ where $x_{p,\tilde{q},X^{}_{{p},\tilde{q}}}$ is the Kähler
coordinate , relative to the complex structure $X^{}_{{p},\tilde{q}}$,  of the Poincare dual to the 3 cycle defining the 4d BH.

The first thing to be noticed is that the Bekenstein-Hawking entropy
for the BH of charge $(p,\tilde{q})$ is just given by the relative normalization of the BH  coherent state with respect to the "vacuum"  $|0,0\ra$.
\be
\frac{\la 0,\l^{-1}_{p,\tilde{q},X^{}_{{p},\tilde{q}}}| 0,\l^{-1}_{p,\tilde{q},X^{}_{{p},\tilde{q}}} \ra_{X^{}_{{p},\tilde{q}},\bar{X}^{}_{{p},\tilde{q}} } }    {\la 0,\l^{-1}_{0,0,X^{}_{{p},\tilde{q}}}=0| 0,\l^{-1}_{0,0,X^{}_{{p},\tilde{q}}}=0 \ra_{X^{}_{{p},\tilde{q}},\bar{X}^{}_{{p},\tilde{q}} }}=\exp \left[ -\frac{1}{\hbar}e^{-K(X_{p,\tilde{q}},\bar{X}_{p,\tilde{q}})}|\l^{-1}_{p,\tilde{q},X_{p,\tilde{q}}}|^2\right]=e^{-S_{BH}(p,\tilde{q})}
\ee
with $\hbar=4/\pi$.

Of course, the relation inside coherent states between $p,\tilde{q}$ and $x,\l^{-1}$ is purely classical, and this is the reason why we are able to reproduce the Bekenstein-Hawking entropy in terms of these coherent states. If instead we want to reproduce the BHW entropy we need to introduce the state $|\psi_{\rm{top}} \ra$, which contains the higher genus corrections. After computing its scalar  product with the BH coherent state
\be
\la\psi_{\rm{top}}| 0,\l^{-1}_{p,\tilde{q},X^{}_{{p},\tilde{q}}} \ra_{X^{}_{{p},\tilde{q}},\bar{X}^{}_{{p},\tilde{q}} {\rm bad}}=\exp \left[ F_{\rm top}(\l_{p,\tilde{q},t^{}_{{p},\tilde{q}},\bar{t}^{}_{{p},\tilde{q}}};t^{}_{{p},\tilde{q}},\bar{t}^{}_{{p},\tilde{q}})  -\frac{1}{\l^2_{p,\tilde{q}}}F_0(t^{}_{{p},\tilde{q}})-\bar{f}_1(\bar{t}_{{p},\tilde{q}})\right]
\ee
we get  the following compact expression for the BHW entropy formula
\be
S_{\rm BHW}(p,q)=\log |\la\psi_{\rm{top}}| 0,\l^{-1}_{p,\tilde{q},X^{}_{{p},\tilde{q}}} \ra_{X^{}_{{p},\tilde{q}},\bar{X}^{}_{{p},\tilde{q}}} |^2 + \frac{1}{\l^2_{p,\tilde{q}}}F_0(t^{}_{{p},\tilde{q}})+\frac{1}{\bar{\l}^2_{p,\tilde{q}}}F_0(\bar{t}^{}_{{p},\tilde{q}})+\frac{\pi}{2}q\phi_{{p},\tilde{q}}
\ee
This representation of the BHW entropy has a very interesting meaning in terms of quantum distribution functions on phase space.
In order to unravel this meaning we need to come back to the discussion on the  normalization of coherent states relative to $|0,0\ra$. Until now we have been considering coherent states defined by acting on the vacuum state by $e^{\alpha a^{+}}$. These coherent states depend holomorphically on 
$(x,\lambda)$, something that was crucial to match the holomorphic anomaly as the corresponding Bogolioubov transform of these states under changes of Kähler polarization. In addition the relative normalization of these states with respect to the ``vacuum'' is different from one and, as we have just shown above, is determined by the Bekenstein-Hawking entropy. A different way to define coherent states is by acting on the ``vacuum'' with the displacement operator in phase space, namely by $\hat{D}(\alpha)|0\ra =
e^{\alpha \hat{a}^{\dagger} -\alpha^{*}\hat{a}}|0\ra$. In this case we get coherent states that are correctly normalized with respect to the ``vacuum'' state.
These coherent states are defined by
\be
| x,\l^{-1} \ra= \exp \left[ -\frac{1}{\hbar} e^{-K} \hat{\bar{\lambda}}^{-1}{\lambda}^{-1} + \frac{1}{\hbar} e^{-K} x^i\hat{\bar{x}}^{\bar{j}} G_{i\bar{j}} +\frac{1}{\hbar} e^{-K} \hat{{\lambda}}^{-1}{\bar{\lambda}}^{-1} - \frac{1}{\hbar} e^{-K} \hat{x}^i{\bar{x}}^{\bar{j}} G_{i\bar{j}}   \right] | 0,0 \ra
\ee
\bea
\label{compl}
\imaginario&=&\int d\mu_{x,\l^{-1}}  | x,\l^{-1} \ra \la \bar{x},\bar{\l}^{-1} |\\
\frac{\la \bar{x}^\prime,\bar{\l}^{-1\prime} | x,\l^{-1} \ra }{\la \bar{0},\bar{0} |0,0 \ra }&=&\exp \left[ -\frac{1}{\hbar} e^{-K} {\bar{\lambda}}^{-1\prime}{\lambda}^{-1} + \frac{1}{\hbar} e^{-K} x^i{\bar{x}}^{\bar{j}\prime} G_{i\bar{j}}   \right] \cdot \\
&&\cdot \exp \left[ +\frac{1}{2\hbar} e^{-K} {\bar{\lambda}}^{-1}{\lambda}^{-1} - \frac{1}{2\hbar} e^{-K} x^i{\bar{x}}^{\bar{j}} G_{i\bar{j}}   \right] \cdot\\
&&\cdot \exp \left[ +\frac{1}{2\hbar} e^{-K} {\bar{\lambda}}^{-1\prime}{\lambda}^{-1\prime} - \frac{1}{2\hbar} e^{-K} x^{i\prime}{\bar{x}}^{\bar{j}\prime} G_{i\bar{j}}   \right]
\eea
Notice that now these coherent states are not holomorphic with respect to $(x,\l^{-1} )$. We will see the rationale of passing from ``bad'' to ``good'' normalized coherent states  in the next subsection. For the time being let us just observe that, by this change of coherent states, we can absorb the $F_{0}$ contribution in the BHW entropy formula. In fact using that
\be
S_{BH}(p,\tilde{q})=\frac{1}{\l^2_{p,\tilde{q}}}F_0(t^{}_{{p},\tilde{q}})+\frac{1}{\bar{\l}^2_{p,\tilde{q}}}F_0(\bar{t}^{}_{{p},\tilde{q}})+\frac{\pi}{2}\tilde{q}\phi_{{p},\tilde{q}}
\ee
we arrive to the result
\be
e^{S_{BHW}(p,q)}=|\la\psi_{\rm{top}}| 0,\l^{-1}_{p,\tilde{q},X^{\rm quan}_{{p},{q}}} \ra_{X^{\rm quan}_{{p},{q}},\bar{X}^{\rm quan}_{{p},{q}} {\rm good}}|^2 \exp \frac{\pi}{2}(q-\tilde{q})\phi^{quan}_{{p},{q}}
\ee
In this representation, the BHW entropy is basically the convolution of the state $|\psi_{\rm top}\ra$ with the BH squeezed state centered in $p,\tilde{q}(p,q)$ with width and squeezed parameters given by $\t$ evaluated at the quantum corrected attractor point. 

This formula is part of what we were looking for, namely to associate the BHW entropy with a concrete quantum distribution function on $H^{3}$ phase space. In fact we will define a BH quantum distribution function by the norm square in the previous formula. Its meaning will be worked in the next subsection.

\subsection{Black Hole Quantum Distribution Function as a Mixed Husimi/anti-Husimi Distribution}

In the last subsection we have expressed the BHW entropy in terms of a black hole quantum distribution function
\be
F^{\rm BH}_{|\psi_{\rm top}\ra}(p,\tilde{q};X_{p,\tilde{q}},\bar{X}_{p,\tilde{q}})=|\la   \psi_{\rm top} |  \l^{-1}_{p,\tilde{q};X_{p,\tilde{q}}} , 0 \ra_{X_{p,\tilde{q}},\bar{X}_{p,\tilde{q}} {\rm good} }|^2
\ee
where $X_{p,\tilde{q}}=X^{\rm quan}_{p,q}$ is the corrected attractor point of the black hole. In order to understand what does this function mean, let us introduce some delta functions
\be
F^{\rm BH}_{|\psi_{\rm top}\ra}=\int d\mu_{x^\prime,\l^{-1\prime}} \delta(x^\prime) \delta(\l^{-1\prime}-\l^{-1}_{p,\tilde{q};X_{p,\tilde{q}}})  
|\la   \psi_{\rm top} |  \l^{-1\prime} , x^\prime \ra_{X_{p,\tilde{q}},\bar{X}_{p,\tilde{q}} {\rm good} }|^2
\ee
If we use the following expressions for the delta functions
\bea
\delta(x^\prime)&=&\int d\mu_{x^{\prime\prime}} \exp \left[ \frac{1}{\hbar} e^{-K}G_{i\bar{j}}x^{i\prime\prime}\bar{x}^{\bar{j}\prime}-
\frac{1}{\hbar} e^{-K}G_{i\bar{j}}x^{i\prime}\bar{x}^{\bar{j}\prime\prime}  \right]\\
\delta(\l^{-1\prime})&=&\int d\mu_{\l^{-1\prime\prime}} \exp \left[
\frac{1}{\hbar} e^{-K} \l^{-1\prime\prime}  \bar{\l}^{-1\prime} -
\frac{1}{\hbar} e^{-K} \l^{-1\prime}  \bar{\l}^{-1\prime\prime}
\right]  
\eea
where $K$ and $G_{i\bar{j}}$ are evaluated at $X_{p,\tilde{q}}$, we obtain
\bea
F^{\rm BH}_{|\psi_{\rm top}\ra}=\int d\mu_{x^{\prime\prime},\l^{-1\prime\prime}}d\mu_{x^\prime,\l^{-1\prime}}
|\la   \psi_{\rm top} |  \l^{-1\prime} , x^\prime \ra_{X_{p,\tilde{q}},\bar{X}_{p,\tilde{q}} {\rm good} }|^2  \nonumber \\
\exp \left[  \frac{1}{\hbar} e^{-K} \l^{-1\prime\prime}  \bar{\l}^{-1\prime}+ 
\frac{1}{\hbar} e^{-K}G_{i\bar{j}}x^{i\prime\prime}\bar{x}^{\bar{j}\prime}
-\frac{1}{\hbar} e^{-K} \l^{-1\prime}  \bar{\l}^{-1\prime\prime}
-\frac{1}{\hbar} e^{-K}G_{i\bar{j}}x^{i\prime}\bar{x}^{\bar{j}\prime\prime}
  \right]  \nonumber\\
\exp  \left[  
-\frac{1}{\hbar} e^{-K} \l^{-1\prime\prime}  \bar{\l}^{-1}_{p,\tilde{q};X_{p,\tilde{q}}}
+ \frac{1}{\hbar} e^{-K} \l^{-1}_{p,\tilde{q};X_{p,\tilde{q}}} \bar{\l}^{-1\prime\prime}
  \right] 
\eea
Taking into account that $|  \l^{-1\prime} , x^\prime \ra$ is eigenstate of $\hat{\l}^{-1},\hat{x}^i$
\bea
&F^{\rm BH}_{|\psi_{\rm top}\ra}=&\int d\mu_{x^{\prime\prime},\l^{-1\prime\prime}}d\mu_{x^\prime,\l^{-1\prime}}
\la   \psi_{\rm top} | e^{-\frac{1}{\hbar} e^{-K} \hat{\l}^{-1}  \bar{\l}^{-1\prime\prime} 
-\frac{1}{\hbar} e^{-K}G_{i\bar{j}}\hat{x}^{i}\bar{x}^{\bar{j}\prime\prime}
}         | \l^{-1\prime} , x^\prime \ra_{X_{p,\tilde{q}},\bar{X}_{p,\tilde{q}} {\rm good} }\nonumber\\& &
_{X_{p,\tilde{q}},\bar{X}_{p,\tilde{q}} {\rm good} }\la \l^{-1\prime} , x^\prime |
e^{  \frac{1}{\hbar} e^{-K} \l^{-1\prime\prime}  \hat{\bar{\l}}^{-1}+ 
\frac{1}{\hbar} e^{-K}G_{i\bar{j}}x^{i\prime\prime}\hat{\bar{x}}^{\bar{j}}    }
|\psi_{\rm top}\ra
  \nonumber\\& &
\exp  \left[  
-\frac{1}{\hbar} e^{-K} \l^{-1\prime\prime}  \bar{\l}^{-1}_{p,\tilde{q};X_{p,\tilde{q}}}
+ \frac{1}{\hbar} e^{-K} \l^{-1}_{p,\tilde{q};X_{p,\tilde{q}}} \bar{\l}^{-1\prime\prime}
  \right] 
\eea
Finally we use the formula (\ref{compl}) and the fact that $x^i_{p,\tilde{q};X_{p,\tilde{q}}}=0$ to obtain
\bea
F^{\rm BH}_{|\psi_{\rm top}\ra}=\int d\mu_{x^{\prime\prime},\l^{-1\prime\prime}}
\la   \psi_{\rm top} | e^{-\frac{1}{\hbar} e^{-K} \hat{\l}^{-1}  \bar{\l}^{-1\prime\prime} 
-\frac{1}{\hbar} e^{-K}G_{i\bar{j}}\hat{x}^{i}\bar{x}^{\bar{j}\prime\prime}
}         
e^{  \frac{1}{\hbar} e^{-K} \l^{-1\prime\prime}  \hat{\bar{\l}}^{-1}+ 
\frac{1}{\hbar} e^{-K}G_{i\bar{j}}x^{i\prime\prime}\hat{\bar{x}}^{\bar{j}}    }
|\psi_{\rm top}\ra
  \nonumber\\
\exp  \left[  \frac{1}{\hbar} e^{-K} \l^{-1}_{p,\tilde{q};X_{p,\tilde{q}}} \bar{\l}^{-1\prime\prime}
+
\frac{1}{\hbar} e^{-K} G_{i\bar{j}}x^{i}_{p,\tilde{q};X_{p,\tilde{q}}} \bar{x}^{\bar{j}\prime\prime}
-\frac{1}{\hbar} e^{-K} \l^{-1\prime\prime}  \bar{\l}^{-1}_{p,\tilde{q};X_{p,\tilde{q}}} 
-\frac{1}{\hbar} e^{-K} G_{i\bar{j}} x^{i\prime\prime}  \bar{x}^{\bar{j}}_{p,\tilde{q};X_{p,\tilde{q}}} 
  \right] \nonumber
\eea
Therefore, the black hole distribution function corresponds to the quantum distribution function at which the map $A(p,q)\to \hat{A}(\hat{p},\hat{q})$ is done by using an operator ordering where $\hat{\l}^{-1}$ and $\hat{x}^i$ go at the front. Since $\hat{\l}^{-1}$ is a creation operator whereas $\hat{x}^i$ is an annihilation operator, this ordering corresponds to a normal ordering with respect to the $\l^{-1}$-sector, but antinormal ordering with respect to the $x$-sector, that is, the black hole distribution function is the mixed Husimi/anti-Husimi quantum distribution function of the state $|\psi_{\rm top}\ra$ \footnote{See appendix A for the definition and properties of Husimi and Wigner quantum distributions.}. For a given state there are several Husimi and anti-Husimi quantum distribution functions, depending on their resolution parameters. The one that corresponds to the black hole distribution function has resolution parameter given by $\t_{IJ}$ evaluated at the corrected attractor point. Notice that, in this derivation, it has been crucial the fact that the coherent states are correctly normalized with respect to $|0,0\ra$.

In summary we have identified the macroscopic BHW entropy as a mixed Husimi/anti Husimi distribution on $H^{3}$. In the next section we will consider from this point of view the microscopic OSV entropy.
\subsection{Ooguri-Strominger-Vafa Conjecture in terms of the Quantization of $H^3$. Physical meaning}

On the basis of the previous results, we can  -assuming the OSV conjecture- represent the black hole mixed ensemble partition function in terms of the BH quantum distribution function
\bea
Z_{\rm BH}(p,\phi)&=&|\la\psi_{\rm{top}}| 0,\l^{-1}_{p,q_\phi,X} \ra_{X,\bar{X} {\rm good}}|^2 \exp -\frac{\pi}{2}q_\phi\phi=\\
&=&F^{\rm BH}_{|\psi_{\rm top}\ra}\left( p,q_\phi;X=p+i\frac{\phi}{2},\bar{X}=p-i\frac{\phi}{2}\right) \exp\left[ +\pi \frac{\phi^I}{2} \I \t_{IJ} \frac{\phi^J}{2}-\frac{\pi}{2}\phi^I \R \t_{IJ} p^J \right]\nonumber
\eea
where $q_{\phi}=-\I \tau \frac{\phi}{2}+\R \t p$ is the charge associated, through the classical attractor equations, to $X=p+i\frac{\phi}{2}$. We can also write
\be
-\frac{\pi}{2}q_{\phi}\phi=+\pi q_\phi (\I\t)^{-1}(q_\phi-\R\t p)
\ee
Therefore, we conclude that the conjectured OSV BH mixed partition function is the mixed Husimi/anti-Husimi quantum distribution function associated with $|\psi_{\rm top}\ra$, with a width given by $(p,\phi)$, times the exponential of the squared of the classical charge $q_\phi$ associated with $(p,\phi)$. In addition, for the conjectured number of microstates we get
\bea
\Omega(p,q)&=&C\int d\phi |\la\psi_{\rm{top}}| 0,\l^{-1}_{p,q_\phi,X} \ra_{X,\bar{X} {\rm good}}|^2 \exp \frac{\pi}{2}(q-{q}_\phi)\phi=\\
&=&C\int d\phi F^{\rm BH}_{|\psi_{\rm top}\ra}\left( p,q_\phi;X=p+i\frac{\phi}{2},\bar{X}=p-i\frac{\phi}{2} \right) \exp\left[ \pi \frac{\phi}{2}\I \t \frac{\phi}{2} + \pi \frac{\phi}{2} (q-\R\t p) \right] \nonumber
\eea
By using the inverse relation we can write the black hole quantum distribution function in terms of $\Omega(p,q)$
\be\label{F}
F^{\rm BH}_{|\psi_{\rm top}\ra}(p,q_\phi;X,\bar{X})
=\int dq^\prime \Omega (p,q^\prime) \exp \left[+\pi (q^\prime-q_\phi)(\I \tau )^{-1} (q_\phi - \R \tau p)  \right]
\ee
The previous formula (\ref{F}), with $\Omega(p,q)$ the number of microstates, is a direct consequence of the OSV conjecture $Z_{BH}=|Z_{top}|^{2}$ and can be used as an alternative form of establishing the conjecture. The integral representation (\ref{F}) of the BH quantum distribution function is here based on the conjecture that the BHW entropy is in fact the Legendre transform of the mixed BH partition function.
However the quantum mechanical meaning of the BH quantum distribution function already leads us, without invoquing any mixed BH ensemble, to an integral representation of the BH quantum distribution function in terms of the Wigner distribution associated with $|\psi_{top}\ra$. This integral representation has a completely different origin to the one derived from the OSV conjecture. However as we will show in a moment in certain limits both integral representations of the BH quantum distribution function formally agree.

In fact we can relate the mixed Husimi/anti-Husimi quantum distribution function with the Wigner function\footnote{This Wigner function $F^W_{|\psi_{\rm top}\ra}(p,q)$ is precisely the one that has been conjectured \cite{Gerasimov:2004yx,Dijkgraaf:2004te,Nekrasov:2004vv} to be equal to the partition function of $V_H$ Hitchin theory \cite{Hitchin:2000sk}. In fact, if one defines the mixed Husimi/anti-Husimi function $F^{nl}_{|\psi_{\rm top}\ra}(p,q)$ corresponding to non-linear polarization $\hat{p}^J=\R \hat{X}^J$, $\hat{q}_I=\R F_I(\hat{X})$ and uses the conjectured representation for $|\psi_{top}\ra$ of \cite{Gerasimov:2004yx,Ooguri:2005vr} in this polarization, one obtains that the expresion of $F^W_{|\psi_{\rm top}\ra}(p,q)$ in terms of $F^{nl}_{|\psi_{\rm top}\ra}(p,q)$ is roughly the partition function of Hitchin theory. This expresion can be obtained, for instance, by inserting $\imaginario=\int d\mu_{X,\bar{X}} |X\ra\la X|$ into (\ref{meanpar}). This conjecture has been studied at 1 loop in \cite{Pestun:2005rp}} associated with $|\psi_{\rm{top}} \ra$ by using the known relation\footnote{See Appendix A}
\bea
F^{\rm BH}_{|\psi_{\rm top}\ra}(p,q_\phi;X,\bar{X})&=&\int dp^\prime dq^\prime F^W_{|\psi_{\rm{top}} \ra}(p^\prime,q^\prime) \\
&&\exp \left[ -\pi \left( (q_\phi-q^\prime) -\R \tau (p-p^\prime)  \right)(\I \tau )^{-1} \left( (q_\phi-q^\prime) -\R \tau (p-p^\prime)  \right) \right. \nonumber\\
&&\left. - \pi (p-p^\prime) \I \t (p-p^\prime)   \right] \nonumber
\eea
Now we can compare this integral representation with the one derived from OSV conjecture. Working for simplicity with $Re \tau=0$, we get 
\bea
F^{\rm BH}_{|\psi_{\rm top}\ra}=&\int dp^\prime dq^\prime &F^W_{|\psi_{\rm{top}}}(p^\prime,q^\prime) \exp \left[ -\pi (q_\phi-q^\prime) (\I\t)^{-1}(q_\phi-q^\prime)-\pi (p-p^\prime) \I \t (p-p^\prime) \right]\nonumber\\
F^{\rm BH}_{|\psi_{\rm top}\ra}=&\int dq^\prime &\Omega(p,q^\prime)\exp \left[ -\pi (q_\phi-q^\prime) (\I\t)^{-1}q_\phi \right]
\eea
In general both relations are different, showing that the OSV degeneracy $\Omega(p,q)$ is not a Wigner distribution function, but a very different object\footnote{In \cite{Ooguri:2004zv} the connection between $\Omega(p,q)$ and  
the Wigner function was done using a topological wave function $\psi_ 
{top}(p+i\phi)$, performing a formal analytic continuation $i\phi =p^\prime$ and interpreting the result as a wave function in real  
polarization. The main problem with these formal manipulations is  
that they ignore the physical meaning of the complex argument of the  
topological wave function as referring to coherent states. In \cite{Verlinde:2004ck}
 the Wigner wave function was actually obtained in the  
limit $\tau\to i\infty$ of an expression that was formally background  
independent. The main problem with this approach is that it is not  
taking into account the non holomorphic corrections to the attractor  
equations.}. Nevertheless, if the black hole is such that the attractor point is located in a region ``at infinity'' at which
\be
\t\to i\infty
\ee
both expressions formally agree. Notice that $F^{W}_{|\psi_{top}\ra}(p,q)$ as well as $\Omega(p,q)$ are  
independent of background, however only $\Omega(p,q)$ has the  
statistical mechanical meaning of microscopic black hole degeneracies.

\section{Discussion}

All the quantum phase space distribution functions are known to contain the same information as the quantum wavefuncion, and we know that choosing a concrete one is a matter of convenience. Although the Wigner function has the good property
\be
\label{goopro}
|\psi(p)|^2=\int dq F^W_{|\psi\ra}(p,q)
\ee  
it fails to be positive. In fact, it oscillates rapidly in regions outside the classical phase space trajectory. On the other hand, Husimi distributions are more frequently used
if the quantum and classical phase space quasi-probability densities themselves are in the focus of attention. This is because, although they fail to have the property (\ref{goopro}), they are obtained by averaging the Wigner function by using gaussians corresponding to the coherent states, in such a way that they are positive. In fact, Wigner is also obtained from anti-Husimi by doing the same averaging process. Since all the distribution functions contain the same information from the point of view of quantum mechanics, we can consider all the information lost in these averaging process to be meaningless\footnote{
This point of view corresponds to taking also into account the proper function kernels.
As explained in appendix A, the distribution function kernel $f$ is trivial only for the Wigner function, but it leaves residual messy ''measures´´ for all other distributions, including the Husimi one. This implies that, if one maintains the Weyl map, expectation values of observables now entail equivalence conversion dressings of the respective kernel functions. As explained in section 0.13 of the first chapter of \cite{Zachos}, the fact that phase-space integrals are complicated by conversion dressing convolutions implies that  
distributions such as $F^H$ cannot be automatically thought of as bona fide probability distributions. Ignoring the equivalence dressings in the 
computation of expectation values results in loss of quantum information effectively coarse-graining to a classical limit. We thank C. Zachos for pointing this fact out to us.}. In our case we have a mixed Husimi/anti-Husimi distribution function and, therefore, it contains in the $\l^{-1}$-sector information that is averaged in the Wigner function, whereas the Wigner function contain in the $x$-sector information that is averaged into the 
black hole distribution function. Nevertheless, since the object
\be
\la \psi_{\rm top} | \l^{-1},x  \ra=\psi_{top}(\l^{-1},x )
\ee
have sense, our mixed Husimi/anti-Husimi function is also positive. Therefore, in some sense, the Wigner function, and $\Omega(p,q)$ in the region $\t\to i\infty$, contain less information in the $\l^{-1}$-sector than the one contained in the BHW entropy, but extra information in the $x$-sector\footnote{Notice also that in the limit $\t\to i\infty$ our Husimi/anti-Husimi distribution function is independent on $q$ and becomes $|\la \psi_{\rm top}|p\ra|^2$. On the other hand, Wigner function is independent of the base point, since it is not constructed by using any coherent state. Therefore, in this limit, what the black hole distribution function is doing is summing the Wigner function values at different values of $q$.}. It is this extra information the one that is contained in OSV conjecture, in the region $\t\to i\infty$, and that it is not contained in BHW formula.


\acknowledgments
We wish to thank P. Resco for the collaboration in the first stage of this work and for  helpful discussions.
S.M. would also like to thank the CERN Theory Group and the High Energy Group of Univ.
Mohammed V (Rabat, Morocco) where part of this work was done for hospitality, specially A. Belhaj and E.H. Saidi.
This work is partially supported by the Spanish DGI contract FPA2003-02877 
and CAM project HEPHACOS P-ESP-00346. 
The work of S.M. is supported by the Spanish Ministerio de Educaci\'on y 
Ciencia through FPU Grant 
AP2002-1386. 

\begin{appendix}
\section{Quantum Phase Space Distribution Functions}
In this appendix we review briefly some of the basic properties of the quantum distribution functions we use along the paper. For a more extensive review on the theory of quantum phase space distribution functions see for instance \cite{qpsdf,Zachos}.
\subsection{Definition}
Let us consider a quantum state $|\psi\ra$. A quantum phase space distribution function $F(p,q)$ associated with this state is a function from which we can obtain the expectation value of every quantum observable $\hat{A}(\hat{p},\hat{q})$ in the state by the averaging formula
\be
\la \psi | \hat{A}(\hat{p},\hat{q}) |\psi\ra=\int dqdp A(p,q) F(p,q)
\ee
where $A(p,q)$ is a classical observable associated with $\hat{A}(\hat{p},\hat{q})$. Due to the fact that $\hat{q}$ and $\hat{p}$ do not commute, there are several ways of doing the map $A(p,q)\to\hat{A}(\hat{p},\hat{q})$ each one corresponding to a different quantum distribution function $F(p,q)$. Each operator ordering give rise to a different kernel function $f(\xi,\eta)$ defined by
\be
e^{\frac{i}{\hbar}(\xi q-\eta p)}\to f(\xi,\eta)e^{\frac{i}{\hbar}(\xi \hat{q}-\eta \hat{p})}
\ee
The scalar function $A^f(p,q)$ associated with $\hat{A}(\hat{p},\hat{q})$ is
\be
A^f(p,q)=\frac{1}{2\pi} \int d\xi d\eta \frac{\tilde{A}(\xi,\eta)}{f(\xi,\eta)} e^{\frac{i}{\hbar}(\xi q-\eta p)}
\ee
where $\tilde{A}(\xi,\eta)$ are the coefficients of the Fourier expansion
\be
\hat{A}(\hat{p},\hat{q})=\frac{1}{2\pi} \int d\xi d\eta \tilde{A}(\xi,\eta) e^{\frac{i}{\hbar}(\xi \hat{q}-\eta \hat{p})}
\ee
and the corresponding quantum distribution function
\be
\label{func}
F^f(p,q)=\frac{1}{4\pi^2} \int d\xi d\eta \la \psi | f(\xi,\eta)e^{\frac{i}{\hbar}(\xi \hat{q}-\eta \hat{p})} |\psi\ra e^{-\frac{i}{\hbar}(\xi q-\eta p)}
\ee

\subsection{Wigner Function}
Wigner function corresponds to Weyl ordering of operators, that is $f^W=1$. This implies that $A^W(p,q)$ is the Weyl transform of the operator $\hat{A}(\hat{p},\hat{q})$
\be
A^W(p,q)=2\int dq^\prime \la q+q^\prime | \hat{A}(\hat{p},\hat{q}) |q-q^\prime \ra e^{-2\frac{i}{\hbar}q\prime p}
\ee
One can write the Wigner function as the Weyl transform of the density operator $|\psi\ra \la \psi |$
\be
F^W(p,q)=\frac{1}{\pi \hbar} \int d q^\prime \la q+q^\prime |\psi\ra \la \psi |q-q^\prime \ra e^{-2\frac{i}{\hbar}q\prime p}
\ee
or, equivalently, as a mean value of the parity operator $\hat{P}$
\be
\label{meanpar}
F^W(p,q)=\frac{1}{\pi \hbar} \la \psi | \hat{D}^\dagger(p,q) \hat{P} \hat{D}(p,q) | \psi \ra
\ee
where $\hat{D}(p,q)=e^{\frac{i}{\hbar}(p \hat{q}-q \hat{p})}$ is the Weyl's unitary displacement operator. From the last expression it is easy to see that $F^W(p,q)\in [-\frac{1}{\pi \hbar},+\frac{1}{\pi \hbar}]$ is a real but non-positive function.

\subsection{Husimi Function}
By using an arbitrary frequency $\omega$ one can define annihilation operators
\be
\hat{a}=\frac{1}{\sqrt{2\hbar\omega}} (\omega \hat{q} + i \hat{p})
\ee
or, more generally, generalized annihilation operators
\be
\hat{b}=\mu \hat{a} + \nu \hat{a}^\dagger
\ee
with $\mu,\nu\in \complejo$, $|\mu|-|\nu|>0$. Their corresponding eigenstates are squeezed states $|\beta\ra$. Husimi distribution functions correspond to anti-normal ordering with respect to this generalized annihilation operators. Therefore
\be
f^H(\xi,\eta)=e^{-\frac{|\beta(\xi,\eta)|^2}{2}}
\ee
where
\be
\beta(\xi,\eta)=\mu \frac{1}{\sqrt{2\hbar\omega}} (\omega \eta + i \xi) + \nu \frac{1}{\sqrt{2\hbar\omega}} (\omega \eta - i \xi)
\ee
Notice that
\be
\hat{D}(p,q)=e^{\beta(p,q) \hat{b}^\dagger-\beta^*(p,q) \hat{b}}
\ee
for every $\omega,\mu,\nu$. Therefore
\be
\label{Husi}
F^H(p,q)=\frac{1}{4\pi^2} \int d\xi d\eta \la \psi |e^{-\beta^*(\xi,\eta)\hat{b}} e^{\beta(\xi,\eta)\hat{b}^\dagger}  |\psi\ra e^{-\frac{i}{\hbar}(\xi q-\eta p)}
\ee
From this expression it is easy derive the relation with $F^W$
\be
F^H(p,q)=\frac{1}{\pi\hbar} \int d q^\prime dp^\prime F^W(p^\prime,q^\prime) e^{-2|\beta(p,q)-\beta(p^\prime,q^\prime)|^2}
\ee
That is, $F^H$ is obtained from $F^W$ by averaging in phase space using gaussians with width and squeezing parameters given by $\omega,\mu,\nu$. This averaging process is similar to the coarse graining effect that is inherent to all experimental measurement processes. By introducing the squeezed states $|\beta\ra$\footnote{Normalized so that $\la \beta^*|\beta\ra =\la 0|0\ra $} into (\ref{Husi}), one can see that this averaging process gives rise to a positive function
\be
F^H(p,q)=\frac{1}{2\pi\hbar} |\la\psi|\beta\ra|^2
\ee

\subsection{Anti-Husimi Function}
Anti-Husimi distribution functions correspond to normal ordering with respect to the generalized annihilation operators
\be
f^{AH}(\xi,\eta)=e^{+\frac{|\beta(\xi,\eta)|^2}{2}}
\ee
\be
\label{AHusi}
F^{AH}(p,q)=\frac{1}{4\pi^2} \int d\xi d\eta \la \psi | e^{\beta(\xi,\eta)\hat{b}^\dagger} e^{-\beta^*(\xi,\eta)\hat{b}} |\psi\ra e^{-\frac{i}{\hbar}(\xi q-\eta p)}
\ee

From the expressions (\ref{AHusi}) and (\ref{func}) particularized for the Wigner function one can derive
\be
F^W(p,q)=\frac{1}{\pi\hbar} \int d q^\prime dp^\prime F^{AH}(p^\prime,q^\prime) e^{-2|\beta(p,q)-\beta(p^\prime,q^\prime)|^2}
\ee
That is, now it is $F^W$ the function that is obtained from $F^{AH}$ by a coarse graining process with width and squeezing parameters given by $\omega,\mu,\nu$. 

\end{appendix}

\bibliographystyle{JHEP-2}

\end{document}